\documentstyle[12pt]{article}
\begin{document}
\baselineskip=14pt

\begin{center}
{\Large {\bf Parity Violation and a Preferred Frame}}

\vskip 0.8cm
            Tsao Chang \\ 
  Center for Space Plasma and Aeronomy Research\\
  University of Alabama in Huntsville\\
          Huntsville, AL 35899\\
      Email: changt@cspar.uah.edu\\

\vskip 0.5cm
\end{center}

\baselineskip=14pt

  Based on parity violation in the weak interaction and evidences 
from neutrino oscillation, a natural choice is that neutrinos 
may be spacelike particles with a tiny mass.   To keep causality
for spacelike particles, a kinematic time under a non-standard 
form of the Lorentz transformation is introduced, which is related 
to a preferred frame.  A Dirac-type equation for spacelike neutrinos 
is further investigated and its solutions are discussed.  This 
equation can be written in two spinor equations coupled together 
via nonzero mass while respecting maximum parity violation. As a
consequence, parity violation implies that the principle of 
relativity is violated in the weak interaction.

\vskip 0.5cm
\noindent
PACS number: 03.65.-w; 14.60.St; 03.30.+p; 

\newpage
\baselineskip=14pt
\noindent
{\bf I. INTRODUCTION}

 Parity violation is a specific feature of the weak interaction. 
It was first argued by T.D. Lee and C.N. Yang in 1956 [1] 
and experimentally established by C.S. Wu in beta-transition 
of polarized Cobalt nuclei[2]. In the standard model, neutrinos 
are naturally massless.  Three flavors of neutrinos are purely 
left-handed, but anti-neutrinos are right-handed. In recent 
years, many convincing evidences for neutrino oscillation 
come from the solar and atmospheric neutrino data have shown that 
neutrinos have tiny mass (about 1 eV) or mass difference [3-5].
  
If neutrino has a tiny rest mass, it would move slower than
light.  When taking a Lorentz boost with a speed
faster than the neutrino, the helicity of the neutrino would
change its sign in the new reference frame.  In another word,
parity would not be violated in the weak interactions.
In order to solve this dilemma, The hypothesis that neutrinos 
may be spacelike particles is further investigated in this 
paper.  To keep causality for spacelike particles, a kinematic 
time under a non-standard form of the Lorentz transformation 
is introduced, which is related to a preferred frame.
   
  Besides neutrino oscillations, the cosmic ray spectrum at 
$E \approx 1-4$ PeV [6] has suggested that the electron neutrino 
is a tachyon.  Using a model to fit data, it yields a value for 
$m^2(\nu_e)\approx -3 \  eV^2$, which is consistent with the 
results from recent measurements in tritium beta decay 
experiments [7-9]. Moreover, the muon neutrino also exhibits a 
negative mass squared [10].

The negative value of the neutrino mass-square simply means:
     $$  E^2/c^2 - p^2 = m_{\nu}^2 c^2 < 0	\eqno  (1)  $$    
The right-hand side in Eq.(1) can be rewritten as ($-m_s^2 c^2$), 
then $m_s$ has a positive value. The subscript $s$ means
spacelike particle, i.e. tachyon.  The negative value on the right
hand side of Eq.(1) means that $p^2$ is greater than $(E/c)^2$.
Based on special relativity and known as re-interpretation rule, 
tachyon as a hypothetical particle was proposed by Bilaniuk et al. 
in the 1960s [11-12]. Some reviews can be found in Ref.[13].

 It is usually to introduce an imaginary mass to consider neutrinos
as tachyons, but these efforts could not reach the point of 
constructing a consistent quantum theory.  Some early investigations
of Dirac-type equations for tachyons are listed in Ref.[14,15].  
An alternative approach was investigated by Chodos et al. [16].  
A form of the lagrangian density for tachyonic 
neutrinos was proposed.  Although they did not obtain a satisfactory 
quantum theory for tachyonic fermions, they suggested that more 
theoretical work would be needed to determine a physically 
acceptable theory. \\ 

\noindent
{\bf II. GGT AND A PREFERRED FRAME }

  To keep causality for spacelike particles, a non-standard form of 
the Lorentz transformation has been studied, which is called the 
generalized Galilean transformation (GGT). The reason is as follows.

Within the context of  the variation principle, It is well known that 
there is the freedom to choose different types of physical time [17].  
In general, the invariant four-line element can be written in arbitrary 
coordinates
   $$ ds^2 =  g_{\mu\nu}dx^\mu dx^\nu  \eqno  	(2) $$
where index $\mu$ (or $\nu$) = 0,1,2,3. The 0th coordinate is the time 
coordinate, which is related to the measured time but may be not 
identical to it.   The action integral for a free particle has a form:
	$$	S = \int(mc)ds = \int L( x^\mu , \dot x^\mu) d\lambda	
\eqno  	(3)  $$
where $\dot x^\mu = dx^\mu/d\lambda$ and $\lambda$ is called the 
evolution parameter, which also plays roll of time.  $L (x^\mu, 
\dot x^\mu)$ is the Lagrangian in four-dimensional space,
  $$  L (x^\mu,\dot x^\mu) = (mc) ds/d\lambda = 
mc(g_{\mu\nu}\dot x^\mu \dot x^\nu)^{1/2} \eqno  (4)$$
When using the Eular-Lagrange equation in terms of the variation 
principle, we can obtain the geodesic equation.  Notice that the above 
Lagrangian is valid for any choice of the evolution parameter $\lambda$
in one reference frame.  This method can be applied to the space-time 
theories in flat space though it is frequently used in general 
relativity for curved space.  Furthermore, when we consider a specific 
evolution parameter as the physical time in all of the inertial 
frames, certain consistent rules must be imposed.  Under this constraint, 
the definitions of physical time have only a few limited choices.

 Besides the definition of relativistic time, another natural time, 
$\lambda = \tilde t$, defined by the generalized Galilean transformation 
(GGT) has been studied [18-21].  Let us start with a simplified 2-D line 
element, $ds^2 = c^2 dT^2- dX^2 $, in a preferred inertial frame $\Sigma
(X, T)$. Considering another inertial frame, $S(x ,t)$, which moves with 
a constant velocity $v < c$ along the $x$ axis respect to $\Sigma$, 
the 2-D form of GGT can be expressed as follows:
   $$     x =  \gamma ( X - vT )  $$
  $$		   \tilde t = \gamma^{-1}T   \eqno  (5)	$$				       
where $\gamma = (1 - v^2/c^2)^{-1/2} $ is the factor of time dilation 
or length contraction.  It is easily seen from Eq.(5) that the 
synchronization of distant events is absolute, independent of the choice 
of the reference frame. It has been shown that GGT is a non-standard 
form of the Lorentz transformation (LT)[18-21].   GGT and LT are 
equivalent if we describe particles with velocity less than (or equal to) 
light.  On the other hand, when describing spacelike particles, GGT 
has its advantages since the GGT time, $\tilde t$, always goes forward.  

In terms of GGT, 4-D line element in Eq.(2) can be written as:
$$   ds^2 = (c d \tilde t - ({\bf v}/c)\cdot d{\bf r})^2 - (d{\bf r})
\cdot d({\bf r})  = (c dt)^2 - (d{\bf r}) \cdot d({\bf r})      
\eqno  (6)  $$
It gives the relationship between GGT time $\tilde t$ and the SR time
$t$. Therefore, GGT time is a supplement to SR time.
For timelike particles, the 4-D momentum can be defined as
	$$	P^\mu = m_o c dx^\mu/ds			\eqno  (7) $$
where the contravariant 4-D momentum $P^\mu = ( {\bf p}, (ds/cd 
{\tilde t})^{-1} m_o c)$.  Using tensor calculus, the covariant 4-D 
momentum can be obtained as $P_{\mu} = g_{\mu\nu} P^{\nu}$.  It can be 
easily proven that the relation of energy and momentum under GGT  is 
the same as in SR [18-21].  For spacelike particles, since $ds^2 < 0$, 
we need to introduce an invariant, $d\tau = \sqrt{(-ds^2)}/c$,  
the contravariant 4-D momentum can be defined as
	$$	P^\mu = m_s dx^\mu/d\tau =m_s \Gamma dx^\mu/d \tilde t = 
(m_s \Gamma {\bf \tilde u_s}, m_s \Gamma c)	\eqno  (8) $$
where $\Gamma = (d\tau /d \tilde t )^{-1}$. In terms of Eq.(8), 
the relation of energy and momentum under GGT for spacelike particles 
is also the same as Eq.(1) [18-21].  As a natural choice,
 we assume that a tachyon has only positive energy in the preferred
frame $\Sigma$. However, the lowest limitation of momentum and energy for 
tachyons are different in a non-preferred frame.  It can be derived 
from Eq.(8) when velocity $\tilde u \rightarrow \infty$:
$$	p_{\infty} = m_s c [1 - ({\bf n} \cdot {\bf v} /c)^2]^{-1/2}
{\bf n}	$$
$$ E_{\infty}  = -m_s c^2 ({\bf u}\cdot {\bf v} /c^2) 
[1 - ({\bf n}\cdot {\bf v} /c)^2]^{-1/2}   \eqno  (9)    $$
where the unit vector ${\bf n} = {\bf \tilde u}/\tilde u$.  In the 
preferred inertial frame, $v$ = 0, then the low limit ${\bf p}_\infty$ 
= $m_s c\bf n$, and $E_\infty$ = 0.   In other inertial frames, the 
lowest energy is not equal to zero, which could have a limited negative 
value. In Eq.(9), when the unit vector $\bf n$ changes its direction 
from  $x$ to $-x$, the sigh of $(\bf u \cdot \bf v)$ is also changed.  
Because neutrinos are created in the weak interaction, the asymmetry
for $E_\infty$ space inversion
may be a source of CP violation in the weak interaction.
If we identify the preferred frame with the cosmic microwave background
radiation (CMBR),  the earth frame has a speed of ($v/c \approx 10^{-3}$) 
with respect to CMBR. Since the mass of e-neutrino is about $1\ eV$, 
we obtain $\Delta E_{\infty}\approx 10^{-3}\ eV $, 
which is a undetectable effect at present time. \\

\noindent
{\bf III. A SPACELIKE DIRAC-TYPE EQUATION}

To follow Dirac's approach [22], the Hamiltonian form of spacelike 
Dirac-type equation for neutrinos can be written in:
$$  \hat E \Psi = c({\vec \alpha} \cdot {\hat p})\Psi + 
           \beta_s m_s c^2 \Psi 	     \eqno (10)  $$
with  ($\hat E = i\hbar \partial /\partial t , {\hat p} = 
-i \hbar \nabla $).  ${\vec \alpha} = (\alpha_1, \alpha_2, 
\alpha_3$) and $\beta_s$ are 4$\times$ 4 matrix, which are defined as
  $$ {\alpha_i} = \left(\matrix{0&\sigma_i\cr
                         \sigma_i&0\cr}\right),  \quad
   \beta_s = \left(\matrix{0&I\cr
                         -I&0\cr}\right)  \eqno (11)   $$
where $\sigma_i$ is 2$\times$2 Pauli matrix, $I$ is 2$\times$2 unit 
matrix.  It is easily to prove that there are commutation 
relations as follows:
 $$ \alpha_i \alpha_j + \alpha_j \alpha_i = 2 \delta_{ij}    $$
      $$	\alpha_i \beta_s + \beta_s \alpha_i =  0  $$
      $$       \beta_s^2 = -1	\eqno    (12)		$$
Furthermore, the relation between 
the matrix $\beta_s$ and the standard matrix $\beta$ is 
$$  \beta_s = \beta \gamma_5 \,,   \quad where \quad
   {\beta} = \left(\matrix{I&0 \cr   0&-I \cr}\right) \,, \quad  
  {\gamma_5} = \left(\matrix{0&I \cr I&0 \cr}\right)
  \eqno	(13)   $$ 
The spacelike Dirac-type equation (10) was briefly studied in [23].
It can be rewritten in covariant forms by multiplying matrics $\beta$ 
and $\gamma_5$.  The covariant form has been discussed in Ref.[16] 
and [25-26] except the sign for the momentum operator can be 
negative.  A more general form of Dirac equation 
with two mass parameters has also been studied in Ref.[27]. 
In addition, Eq.(10) was investigated in a different way [28].

 We now study the spin-$\frac{1}{2}$ property of the neutrino (or
antineutrino) as a tachyonic fermion.  Denote the wave function 
$\Psi$ by two spinor functions $\varphi ({\vec x},t), 
\chi ({\vec x},t)$ the spacelike Dirac-type equation (10) can be 
rewritten as a pair of two-component equations:
$$ i\hbar \frac{\partial \varphi}{\partial t} = -ic 
    \hbar {\vec \sigma} \cdot  \nabla \chi + m_s c^2 \chi   $$
	$$ i\hbar \frac{\partial \chi}{\partial t} = -ic \hbar 
\vec{\sigma} \cdot \nabla \varphi - m_s c^2 \varphi  
\eqno (14)   $$
From Eq.(14), the conserved current can be derived:
$$  \frac{\partial \rho}{\partial t} +
           \nabla \cdot {\vec j} = 0  \eqno (15)   $$
and we obtain
  $$ \rho = \Psi^{\dag}  \gamma_5 \Psi ,  \quad
  {\vec j} = c(\Psi^{\dag} \gamma_5 {\vec \alpha} \Psi) 
     \eqno (16)  $$
where $\rho$ and $\vec j$ are probability density and current; 
$\Psi^{\dag}$ is the Hermitian adjoint of $\Psi$ .

Considering a plane wave along the $z$ axis for a right-handed 
particle, $\bar \nu$, the helicity $H = ({\vec \sigma} \cdot 
{\vec p})/|{\vec p}| = 1 $, then Eq.(14) yields the following 
relation:
  $$	 \chi = \frac{cp - m_s c^2}{E} \varphi   \eqno	(17) $$

The plane wave can be represented by
$ \Psi(z,t)=\psi_{\sigma}exp[\frac{i}{\hbar}(pz-Et)]$,
where $\psi_{\sigma}$ is a four-component bispinor.  Substituting
this bispinor into the wave equation (10) or (14), the explicit form
of two bispinors with positive-energy states are listed as follows:
$$ \psi_1=\psi_{\uparrow (+)} = N \left(\matrix{1\cr
     0\cr A \cr 0 \cr}\right), \quad
    \psi_2= \psi_{\downarrow (+)}  = N \left(\matrix{0\cr
                       -A \cr 0 \cr 1 \cr}\right)    \eqno (18) $$
and other two bispinors with the negative-energy states are:
$$ \psi_3=\psi_{\uparrow (-)} = N \left(\matrix{1\cr
     0 \cr -A \cr 0 \cr}\right), \quad
    \psi_4= \psi_{\downarrow (-)}  = N \left(\matrix{0\cr
                       A \cr 0 \cr 1 \cr}\right)    \eqno (19) $$
where the component $A$ and the normalization factor $N$ are
$$  A=\frac{cp-m_s c^2}{|E|},  \quad
    N=\sqrt{\frac{p+m_s c}{2m_s c}}        \eqno (20)  $$
For $ \psi_1=\psi_{\uparrow (+)}$, the conserved current in Eq.(15)
becomes:
$$ \rho = \Psi_1{^\dag} \gamma_5 \Psi_1=\frac{|E|}{m_s c^2}, \quad
     j = \frac{p}{m_s}     \eqno (21)  $$
Clearly, the ratio $j/\rho$ represents the superluminal speed $u_s$.
For $ \psi_2=\psi_{\downarrow (+)}$, the density $\rho$ is negative
so that it should be discarded.  If we consider the negative states 
as mathematics solutions in the preferred frame, then $\psi_1 = 
\psi_{\uparrow (+)}$ is the only solution with physical identity
i.e. ${\bar \nu}_R$.
It gives a natural choice that antineutrino is right-handed only.  
Therefore, maximum parity violation implies that the principle of 
relativity is violated in the weak interaction.  If we identify 
the preferred frame with CMBR, the earth frame can be considered 
as the preferred frame approximately ($v/c \approx 10^{-3}$). 
Further study on the negative states and negative $\rho$ may be 
associated with complicated mathematics under GGT (see [24,25]), 
which will not be discussed here. 

Let ${\bar\Psi} = \Psi^{\dag} \beta $, we can obtain the following 
scalars:
$$  {\bar\Psi_1} \Psi_1  = {\bar\Psi_3} \Psi_3  = 1 
   \eqno (22)  $$
$$  {\bar\Psi_2} \Psi_2  = {\bar\Psi_4} \Psi_4  = -1 
   \eqno (23)  $$
In addition, the pseudo scalar for each spinor satisfies:
$$  {\bar\Psi} \gamma_5 \Psi  = 0   \eqno (24)  $$
 
\noindent
{\bf IV.  PARITY VIOLATION FOR NEUTRINOS}
  
In order to compare the spacelike Dirac-type equation Eq.(10) with
the two component Weyl equation in the massless limit, 
we now consider a unitary transformation of $\varphi$ and $\chi$ :
  $$   \xi= {1 \over {\sqrt 2}} (\varphi + \chi)  , \quad
      \eta = {1 \over {\sqrt 2}} (\varphi - \chi)  \eqno (25)  $$
where $\xi ({\vec x},t)$ and $\eta ({\vec x},t)$ are two-component 
spinor functions. In terms of $\xi$ and $\eta$, Eq.(16) becomes 
  $$ \rho = \xi^{\dag} \xi - \eta^{\dag} \eta ,  \quad
 {\vec j} = c(\xi^{\dag} {\vec\sigma} \xi + \eta^{\dag}{\vec \sigma}
 \eta) \eqno (26)  $$
Moreover, Eq.(10) can be rewritten in Weyl representation: 
$$ i\hbar \frac{\partial \xi}{\partial t} = -ic \hbar {\vec \sigma}
    \cdot \nabla \xi - m_s c^2 \eta   $$
$$ i\hbar \frac{\partial \eta}{\partial t} = ic \hbar {\vec \sigma}
     \cdot  \nabla \eta + m_s c^2 \xi  \eqno (27)   $$
In the above equations, both $\xi$ and $\eta$ are coupled via 
the mass term $m_s$.

Comparing Eq.(27) with the well known Weyl equation, we take a 
limit, $m_s = 0$, then the first equation in Eq.(27) reduces to
	$$   \frac{\partial \xi_{\bar{\nu}}}{\partial t} = -c
 {\vec \sigma} \cdot \nabla \xi_{\bar{\nu}}     \eqno (28)   $$
In addition, the second equation in Eq.(27) vanishes because 
$\varphi = \chi$ when $m_s = 0$.

Eq.(28) is the two-component Weyl equation for describing 
anti-neutrinos ${\bar{\nu}}$, which is related to the maximum parity 
violation discovered in 1956 by Lee, Yang and Wu [1,2].  They pointed 
out that no experiment had shown parity to be a good symmetry
 for weak interactions.  Now we see that, in terms of Eq.(27), once 
if neutrino has some mass, no matter how small it is, two equations 
should be coupled together via the mass term while still respecting 
maximum parity violation. 

 Indeed, the Weyl equation (28) is only valid for the antineutrinos 
since the antineutrino always has right-handed spin, which is 
opposite to the neutrino.  
 In order to describe the left-handed neutrino, we now take a 
minus sign for the momentum operator, then Eq.(2) becomes
$$  \hat E \Psi_\nu = -c ({\vec \alpha} \cdot {\hat p})\Psi_\nu + 
           \beta_s m_s c^2 \Psi_\nu 	     \eqno (29)  $$
Similar to the solutions for Eq.(10), Eq.(29) yields one physical 
solution for the neutrino: $\psi_\nu = \psi_{\downarrow (+)}$.
Therefore, only $\bar \nu_R$ and $\nu_L$ exist in nature.
From Eq.(29), the two-component Weyl equation for massless 
neutrino becomes: 
 $$   \frac{\partial \xi_\nu}{\partial t} = c{\vec \sigma} \cdot
           \nabla \xi_\nu      \eqno (30)   $$
 Some related discussions the CPT theorem can be found in Ref.[29].\\  

\noindent
{\bf V. REMARKS}

 Based on parity violation in the weak interaction and evidences 
from neutrino oscillation, a natural choice is that neutrinos 
may be spacelike particles with a tiny mass.  A Dirac-type equation 
for spacelike neutrinos is further investigated and its solutions 
are discussed.  This equation can be written in two spinor equations 
coupled together via nonzero mass while respecting maximum parity 
violation.  As a consequence, parity violation implies that the 
principle of relativity is violated in the weak interaction.

   Spacelike neutrinos have many peculiar features, which 
are very different from all other particles.  Neutrino has 
left-handed spin in any reference frame, but  
 anti-neutrino always has right-handed spin.  This means 
that the speed of neutrinos must be equal to or greater than 
the speed of light. Otherwise, the spin direction of neutrino would 
be changed in some reference frames when taking a Lorentz boost.
Moreover, as shown in this paper, the energy of a tachyonic 
neutrino (or anti-neutrino), $E_\nu$, could be negative in 
non-preferred frames, which was studied in Ref.[30].

 The electron neutrino and the muon neutrino may have slightly
different proper masses.  It provides a natural explanation why 
the numbers of e-lepton and $\mu$-lepton are conserved respectively
in low energy experiments.

 Comparing with the electron mass, the mass term of the e-neutrino
in Eq.(10) is approximately close to zero.  Moreover, from Eq.(24),
${\bar\Psi} \gamma_5 \Psi  = 0$ for Spacelike neutrinos.  It means
that the mass term in Eq.(10) can be negligible in most experiments.  
In fact, the momentum of a neutrino is much greater than ($m_s c$) in 
most measurements. For instance, let $p_s = 10(m_s c) = 16 \ eV/c$,
the speed of the e-neutrino is about $u_s = 1.005 c$. It also yields 
the component of bispinor $A \simeq 1$ in Eq.(18). Therefore, 
spacelike neutrinos behave just like the massless neutrinos. This  
similarity may also play role at the level of  SU(2) gauge theory.

 According to special relativity [31], if there is a spacelike 
particle, it might travel backward in time.  Besides the re-
interpretation rule introduced in the 1960s, another approach is to 
introduce a kinematic time under GGT [18-21].  Therefore, special 
relativity can be extended to the spacelike region without 
causality violation.\\

\vskip 0.4cm
 The author is grateful to G-j. Ni for helpful discussions. Thanks 
also due to helpful correspondence with E. Recami and H-B. Ai. \\

\end{document}